\documentclass[aps,prb,twocolumn,amsmath,showpacs,amssymb,superscriptaddress]{revtex4}
\usepackage{amsmath}
\usepackage{graphicx}
\usepackage{amssymb}

\newcommand{\pdag}{{\phantom{\dagger}}}
\usepackage{dsfont}
\usepackage{color}

\begin{document}

\title{Renormalization of anticrossings in interacting quantum wires with Rashba and Dresselhaus spin-orbit couplings}

\author{Tobias Meng}
\affiliation{Department of Physics, University of Basel, Klingelbergstrasse 82, CH-4056 Basel, Switzerland}
\author{Jelena Klinovaja}
\affiliation{Department of Physics, Harvard University, Cambridge, Massachusetts 02138, USA}
\author{Daniel Loss}
\affiliation{Department of Physics, University of Basel, Klingelbergstrasse 82, CH-4056 Basel, Switzerland}

\begin{abstract}
We discuss how electron-electron interactions renormalize the spin-orbit induced anticrossings between different subbands in ballistic quantum wires. Depending on the ratio of spin-orbit coupling and subband spacing, electron-electron interactions can either increase or decrease
 anticrossing gaps. When the anticrossings are closing due to a special combination of Rashba and Dresselhaus spin-orbit couplings, their gap approaches zero as an interaction dependent power law of the spin-orbit couplings, which is a consequence of Luttinger liquid physics. Monitoring the closing of the anticrossings allows to directly measure the related renormalization group  scaling dimension in an experiment. If a magnetic field is applied parallel to the spin-orbit coupling direction, the anticrossings experience different renormalizations. Since this difference is entirely rooted in electron-electron interactions, unequally large anticrossings also serve as a direct signature of Luttinger liquid physics. Electron-electron interactions furthermore increase the sensitivity of conductance measurements to the presence of anticrossing.
\end{abstract}

\date{\today}

\maketitle

\section{Introduction}
The presence of the electronic spin degree of freedom in nanoscale and mesoscale semiconductor systems has driven a large amount of research during the past decades. Prominent examples range from spintronics,\cite{spintronics_rev} over spin qubits,\cite{loss_vinc,qbits_rev} to topological insulators, and the quantum spin Hall effect.\cite{ti,qshe1,qshe2,qshe3} One important effect is thereby the coupling between the electron spin and its motion. In semiconductor systems, this spin-orbit coupling is usually distinguished into the Rashba coupling,\cite{rashba} generated by structural inversion symmetry breaking, and the Dresselhaus coupling\cite{dresselhaus} due to bulk inversion symmetry breaking. In some one-dimensional systems, such as carbon nanotubes, spin-orbit coupling can lead to helical modes, modes in which opposite spins propagate in opposite directions. \cite{Klinovaja_CNT} In other systems, such as quantum wires and nanoribbons, only the combination of  spin-orbit coupling and an externally applied magnetic field allows for the generation of helical\cite{streda_03,charis_gaas_soi_gap_10,helical_bilayer,helical_graphene,helical_mos2} and fractional helical\cite{oreg_fract,meng_fract} phases. In conjunction with proximity-induced superconductivity, these systems have been proposed to host Majorana zero modes \cite{lutchyn_10,oreg_10} and parafermionic bound states, \cite{oreg_fract,klinovaja_pf_rashba} respectively.

In quantum wires with multiple subbands, the spin-orbit coupling gives rise to both intra- and inter-subband couplings. While the intra-subband spin-orbit coupling results in the usual relative shift of the dispersions for spin up and spin down in momentum space, the inter-subband spin-orbit coupling lifts crossings of different bands of opposite spin,\cite{moroz_barnes_99,governale_02,egues_02,egues_03,Daniel_anticrossings,mireles_01,Daniel_SOI_interband} much like the magnetic field does in the single subband case.\cite{streda_03} The lifting of the inter-subband crossings furthermore leads to a regime of (partial) spin polarization of the current inside the wire,\cite{governale_02,perroni_07,Daniel_anticrossings,Daniel_SOI_interband} which resembles the quasi-helical regime in single subband wires. In a bosonized language, the partially gapped quasi-helical regime in single subband Rashba nanowires can be understood as a consequence of the magnetic field being a relevant perturbation in the renormalization group (RG) sense with respect to the gapless quantum wire fixed point. During the RG flow, electron-electron interactions renormalize the magnetic field to stronger values.\cite{braunecker_prb_10} In the present work, we analyze to what extent electron-electron interactions renormalize the analogous inter-subband anticrossing gaps in multi-subband quantum wires with spin-orbit couplings. We find that depending on system parameters, the inter-subband anticrossing gaps either grow or shrink in the presence of electron-electron interactions. 

The paper is organized as follows. First, we define a two subband model in Sec.~\ref{sec:model}, and perform an RG analysis in Sec.~\ref{sec:rg}. Section~\ref{sec:closing} 
discusses how the monitoring of anticrossings as a function of an applied electric field allows to measure Luttinger liquid power laws. In Sec.~\ref{sec:outlook}, we give an outlook on anticrossings occurring at the bottom of the second subbands, which are not captured by our Luttinger liquid approach. We close by commenting on the effects of an external magnetic field, including its interplay with the anticrossings, in Sec.~\ref{sec:outlook2}, and discuss the sensitivity of transport measurements to the presence of the anticrossings in an interacting wire in Sec.~\ref{sec:trans}.

\section{Experimental setup and model}\label{sec:model}
\begin{figure}
\centering
\includegraphics[scale=0.4]{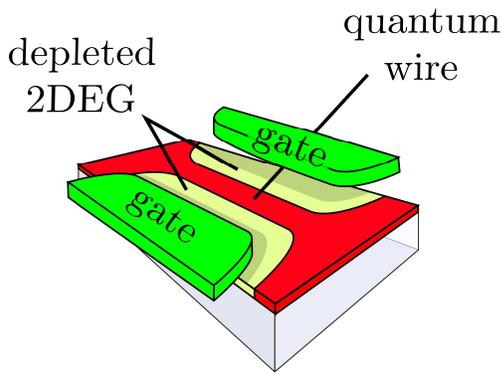}\quad
\caption{The proposed experimental setup consists of a two-dimensional electron gas (2DEG), in which the electrons are confined to a quasi one-dimensional wire region (red). The confinement is due to electrostatic gates placed atop the 2DEG (green), and depleting it outside the wire region (light yellow). By modulating the electrostatic potential of the gates, the width of the wire can be changed in-situ. The electron density inside the wire can be controlled by an additional gate (not shown).}
\label{fig:setup_2deg}
\end{figure}

In this work we focus on ballistic quantum wires defined in two-dimensional electron gases (2DEGs) by means of electrostatic gates since these systems offer a large in situ tunability  (see Fig.~\ref{fig:setup_2deg}). Our analysis is, however, equally applicable to other types of quantum wires with spin-orbit couplings. Together with a gate modulating the electron density in the wire, the setup depicted in Fig.~\ref{fig:setup_2deg} allows one to control the number of occupied subbands, their fillings, and the energy difference between the band bottoms. Concretely, we consider 2DEGs defined in InAs heterostructures because of their relatively large spin-orbit coupling. Of particular interest are samples exhibiting both sizable Rashba and Dresselhaus spin-orbit couplings. While the latter is fixed for a given sample, the Rashba spin-orbit coupling can be controlled by changing an applied electric field.\cite{rashba_control_1,rashba_control_2} In particular, this allows us to access the regime of partial compensation between the two couplings\cite{loss_equal, bernevig_equal} discussed in Sec.~\ref{sec:closing}.

Using units of $\hbar=1$, we model this experimental setup by the Hamiltonian
\begin{align}
H &= H_{\rm 2D}+H_{\rm int.} + H_{\rm R} + H_{\rm D} + H_{\rm conf.}\label{eq:Ham}\, ,\\
 H_{\rm 2D} &= \int d^2r\,\Psi^{\dagger}(\boldsymbol{r})\,\left(\frac{-\partial_x^2-\partial_y^2}{2m}-\mu\right)\,\mathds{1}_{2\times2}\,\Psi(\boldsymbol{r})~,\nonumber\\
 H_{\rm int.}&=\int d^2r\,\int d^2r'\,U(\boldsymbol{r}-\boldsymbol{r}')\,\rho(\boldsymbol{r})\,\rho(\boldsymbol{r}')~~,\nonumber\\
H_{\rm R} &=\alpha\,\int d^2r \,\Psi^{\dagger}(\boldsymbol{r})\,\left(i\sigma_x \partial_y-i\sigma_y\partial_x\right)\,\Psi(\boldsymbol{r})\nonumber~,\\
H_{\rm D} &=  \beta\,\int d^2r \,\Psi^{\dagger}(\boldsymbol{r})\,\left(i\sigma_x \partial_x-i\sigma_y\partial_y\right)\,\Psi(\boldsymbol{r})~,\nonumber
\end{align}
where $\Psi(\boldsymbol{r}) = (c_{\uparrow}(x,y), c_{\downarrow}(x,y))^T$ is the vector of annihilation operators $c_{\nu}(x,y)$ for electrons of spin $\nu$ at position $\boldsymbol{r}=(x,y)^T$. The electrons have an effective mass $m$, and their chemical potential is denoted by $\mu$. The total charge densities $\rho(\boldsymbol{r}) = \sum_{\nu} c_{\nu}^\dagger(\boldsymbol{r}) c_{\nu}^\pdag(\boldsymbol{r})$ interact via the (screened) Coulomb repulsion $U(\boldsymbol{r}-\boldsymbol{r}')$, $\alpha$ denotes the Rashba and $\beta$  the Dresselhaus spin-orbit coupling strength, and $\sigma_i$ are the Pauli matrices. For a heterostructre that has been grown along the crystallographic [001] direction, the $x$ and $y$ directions correspond to the crystallographic [100] and [010] directions, respectively.\cite{winkler_book}

To make the possible partial compensation of Rashba and Dresselhaus spin-orbit couplings more apparent, it is useful to perform a change of coordinates both in real space and spin space by introducing $x'= -(x-y)/\sqrt{2}$, $y'=-(y+x)/\sqrt{2}$, $\sigma_{x'}= (\sigma_{x}-\sigma_{y})/\sqrt{2}$, and $\sigma_{y'}=(\sigma_{y}+\sigma_{x})/\sqrt{2}$. This transformation yields
\begin{align}
H_{\rm SOI} =& H_{\rm R} + H_{\rm D}\nonumber\\
=&-(\alpha+\beta)\,\int d^2r' \,\Psi^{\dagger}(\boldsymbol{r}')\,i\,\sigma_{x'} \partial_{y'}\,\Psi(\boldsymbol{r}')\label{eq:soi}\\
&+(\alpha-\beta)\,\int d^2r' \,\Psi^{\dagger}(\boldsymbol{r}')\,i\,\sigma_{y'}\partial_{x'}\,\Psi(\boldsymbol{r}')\nonumber~.
\end{align}
The first term in Eq.~\eqref{eq:soi}, proportional to $(\alpha+\beta)$, couples different subbands due to its derivative in the $y'$-direction, while the second term, proportional to $ (\alpha-\beta)$, corresponds to a spin-orbit coupling within a given subband. This latter term could be removed from the Hamiltonian by virtue of the gauge transformation $c_{\sigma}(x',y') = e^{i\sigma_zx'k_{\rm SO}}\,c_{\sigma}'(x',y')$ at the expense of introducing an additional phase factor for the inter-subband term.\cite{braunecker_prb_10} In spite of simplicity of this gauge transformation, we do not use it in the present work. The transversal confinement of the electrons due to electrostatic gates, finally, is modeled by a harmonic potential. We focus on a confinement parallel to the $x'$-axis, which is described by the Hamiltonian
\begin{align}
H_{\rm conf.} = \int d^2r'\,\Psi^{\dagger}(\boldsymbol{r}')\,\frac{1}{2}m\Omega^2y'{}^2\,\mathds{1}_{2\times2}\,\Psi(\boldsymbol{r}'). \label{eq:conf}
\end{align}
This potential gives rise to electronic subbands whose transversal wave functions are harmonic oscillator eigenstates. The bottoms of the subbands are offset by an energy difference of $\delta_{12} = \Omega$.

\subsection{Luttinger liquid description}
For simplicity, we focus on the case in which only the lowest two subbands are occupied, and neglect all higher subbands. The effects of higher subbands are similar to the mixing observed in the two subband model,\cite{moroz_barnes_99,mireles_01,governale_02,egues_02,egues_03,liu_09,Daniel_SOI_interband,Daniel_anticrossings} and their neglect restricts our model to sufficiently low (Fermi) energies, such that the coupling to the third subband is not yet important. 

In order to analyze the effects of Coulomb repulsion, we derive a Luttinger liquid description of the system by first decomposing the two-dimensional fermionic operators into operators acting within the different subbands $n=1,2$; $c_{\sigma}(x',y')\to c_{n,\sigma}(x')$. After a rotation in spin space, which brings the intra-subband spin-orbit coupling to a diagonal form, the kinetic energy and intra-subband spin-orbit couplings are described by the Hamiltonian
\begin{align}
H_0&=\int dx'\sum_{\sigma}c_{1,\sigma}^\dagger(x') \left(\frac{-\partial_{x'}^2}{2m}-\mu\right)c_{1,\sigma}^\pdag(x')\nonumber\\
&+\int dx'\sum_{\sigma}c_{2,\sigma}^\dagger(x') \left(\frac{-\partial_{x'}^2}{2m}+\delta_{12}-\mu\right)c_{2,\sigma}^\pdag(x')\label{eq:2band_ham_simpl}\\
&+\int dx'\sum_{n,\sigma}\,(\alpha-\beta)\,\sigma\,c_{n,\sigma}^{\dagger}(x')\,i\partial_{x'}\,c_{n,\sigma}(x')\nonumber~,
\end{align}
where $\sigma =\uparrow,\downarrow\equiv+1,-1$. Next, we restrict the Hamiltonian to low energy excitations close to the Fermi momenta $\pm k_{Fn}$ in the lowest two subbands, which gives rise to left-moving and right-moving modes, $c_{n,\sigma}(x') \approx e^{ix'k_{Fn}}R_{n,\sigma}(x')+e^{-ixk_{Fn}}L_{n,\sigma}(x')$, and linearize the kinetic energy around these momenta. Here, $k_{Fn}$ denotes the Fermi momentum in the $n$th subband. In this effective low energy description, the screened Coulomb interaction yields a number of matrix elements connecting the low energy modes. For our purpose, however, only the leading order terms corresponding to (approximately) zero momentum transfer need to be kept track of. In particular, we have checked that additional sine-Gordon terms corresponding to, for instance, backscattering processes, are RG irrelevant, or less relevant than and competing with the sine-Gordon terms corresponding to inter-subband spin-orbit coupling. The Coulomb repulsion can thus be described by renormalized values of the velocities and Luttinger liquid parameters in the charge sectors of each of the bands, along with a term coupling the charge densities in the two bands.\cite{meng_2bands_13} Bosonizing the effective low energy Hamiltonian,\cite{giamarchi_book} we obtain the Luttinger liquid description
\begin{align}
H = H_{1} + H_2 + H_{12}^{\rm Coulomb} + H_{12}^{\rm SOI}~,\label{eq:ll_ham0}
\end{align}
where
\begin{align}
H_i &= \int \frac{dx'}{2\pi}\,\sum_{j=c,s}\left[\frac{u_{ij}}{K_{ij}}(\partial_{x'}\phi_{ij})^2+u_{ij}K_{ij}(\partial_{x'}\theta_{ij})^2\right] \label{eq:ll_ham}
\end{align}
describes the bosonic charge $c$ and spin $s$ excitations in band one and two, propagating with effective velocities $u_{ij}$, and characterized by Luttinger liquid parameters $K_{ij}$. In our approximation, the velocities and Luttinger liquid parameters are given by $u_{ij} = v_{Fi}/K_{ij}$, $K_{ic} = [1+2U/(\pi v_{Fi})]^{-1/2}$, and $K_{is}=1$ ($U$ is the zero momentum transfer Coulomb matrix element, and $v_{Fi}$ denotes the Fermi velocity in band $i$). The fields $\phi_{ij}$ relate to the integrated charge and spin densities in band $j$, while the fields $\theta_{ij}$ are proportional to the integrated charge and spin currents.\cite{giamarchi_book} They obey the canonical commutation relations $[\phi_{ij} (x), \theta_{i'j'}(x') ]=\delta_{ii'}\delta_{jj'}(i\pi/2) {\rm sgn}(x'-x)$. The Luttinger liquid theory is valid at length scales larger than a short-distance cutoff $a$. We choose the bare value of this cutoff to be given by the lattice constant $a_0$. 

The Coulomb repulsion between the bands yields the term
\begin{align}
H_{12}^{\rm Coulomb} = \int \frac{dx'}{2\pi}\, 2U_{12}\,(\partial_{x'}\phi_{1c})(\partial_{x'}\phi_{2c})~,
\end{align}
with $U_{12} = 2U/\pi$.

\begin{figure}
\centering
\includegraphics[scale=0.85]{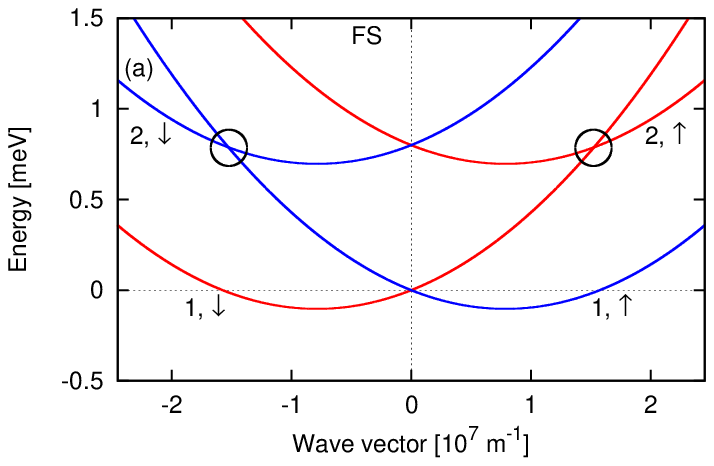}\\
\includegraphics[scale=0.85]{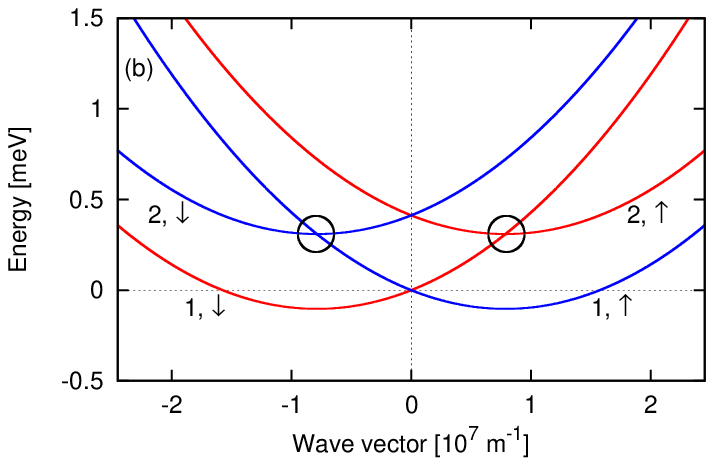}\\
\includegraphics[scale=0.85]{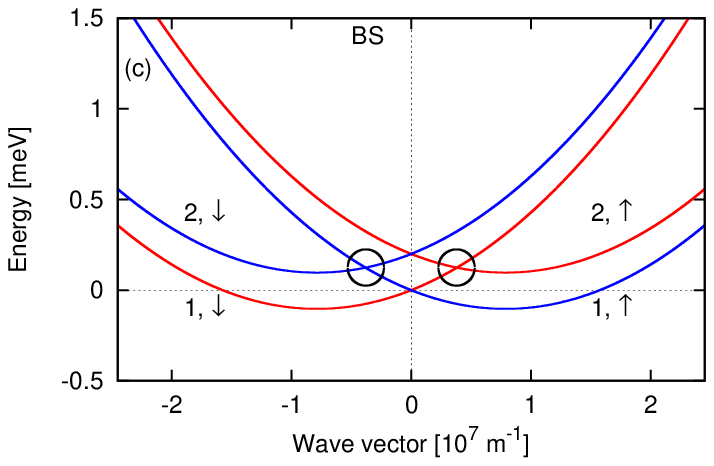}
\caption{Bandstructure of the two-subband quantum wire in the non-interacting case and without interband couplings (the labels $i,\sigma$ indicate subband index and spin polarization). The spectra are calculated using Eqs.~\eqref{eq:Ham} and \eqref{eq:soi} after setting $U = 0$, $\alpha+\beta = 0$, and $\alpha-\beta = \alpha_0$. The red circles highlight the inter-subband crossings that are lifted in the presence of inter-subband spin-orbit coupling. $(a)$ Band structure calculated for a large energy difference of $\delta_{12} = 0.8\,\text{meV}$ between the bottoms of the first and second subbands, corresponding to a situation in which the intersubband spin-orbit coupling induces forward scattering (FS). $(b)$ The case where the crossings occur at the bottoms of the second bands (with an offset of $\delta_{12} \approx 0.4\,\text{meV}$).  Finally, $(c)$ indicates that mall band offsets, here $\delta_{12} = 0.2\,\text{meV}$, lead to interband backscattering (BS). Higher unoccupied bands are neglected.} 
\label{fig:scatterings}
\end{figure} 

The inter-subband spin-orbit coupling $H_{12}^{\rm SOI}$, finally, lifts the crossings between subbands which would be present in the bandstructure otherwise, see Fig.~\ref{fig:scatterings}. For this plot and in the remainder, we use the the effective mass $m = 0.0229 \, m_e$ in terms of the electron rest mass $m_e$, the InAs lattice constant $a_0 = 6.0583\, \text{\AA}$, and we consider a bare intra-subband spin-orbit coupling $\alpha-\beta=\alpha_0 = k_{SO}/m$ corresponding to a spin-orbit length of $k_{SO}^{-1} =127\,\text{nm}$.\cite{winkler_book, fasth07} The value of the Coulomb matrix element $U$ depends on microscopic details, and in particular on the screening length. We choose it such that the Luttinger liquid parameter in the first subband takes the experimentally realistic value\cite{spectroscopy_02,auslaender_spin_charge_seperation_exp_05,jompol_kc_ks} $K_{1c} = 0.65$ when the chemical potential matches the energy of the crossing between the lower subbands $1,\uparrow$ and $1,\downarrow$ (in Fig.~\ref{fig:scatterings}, this corresponds to $\mu=0$).

In the Luttinger liquid picture, the inter-subband spin-orbit coupling corresponds to a sine-Gordon term. This term, in general, is rapidly oscillating in the space due to the finite momentum transfer in the scattering process and can, thus, be neglected.\cite{giamarchi_book} Only when the chemical potential is tuned to (close to) the subband crossing points, as shown in
Fig.~\ref{fig:scatterings}, the sine-Gordon terms are non-oscillating (very slowly oscillating), and need to be kept. In this case, the inter-subband spin-orbit coupling corresponds to
\begin{align}
H_{12}^{\rm SOI} &=\int\frac{dx'}{2\pi a}\,\left(\alpha_{12}^{(1)}\,\cos\left(\Psi_1^{f,b}\right)+\alpha_{12}^{(2)}\,\cos\left(\Psi_2^{f,b}\right)\right)~,\label{eq:hsoi}
\end{align}
where we have dropped the Klein factors\cite{giamarchi_book}, and where $\alpha_{12}^{(1)} = \alpha_{12}^{(2)} =  (\alpha+\beta)\sqrt{m\delta_{12}/2}$. The arguments of the sine-Gordon term $H_{12}^{\rm SOI}$ depend on whether the (eventually lifted) band crossing points occur between modes moving in the {same} or in opposite directions. In the former case, $H_{12}^{\rm SOI}$ encodes inter-subband {forward scattering} (indicated in Eq.~\eqref{eq:hsoi} by the superscript $f$), see Fig.~\ref{fig:scatterings}$(a)$, while it corresponds to inter-subband {backscattering} otherwise (superscript $b$), see Fig.~\ref{fig:scatterings}$(c)$. The interesting intermediate case, in which the crossing occurs at the bottom of the upper bands as shown in  Fig.~\ref{fig:scatterings}$(b)$, is not accessible in our bosonized calculation. If the chemical potential is close to the band bottom, the state of the system is sensitive to the band curvature which we neglect, however. The electrons are furthermore strongly influenced by electron-electron interactions and interband pair tunneling processes,\cite{meng_10} which may induce a partial gap in the system.\cite{meyer_07} We therefore expect our analysis to hold only as long as the Fermi velocities in the second band are sufficiently large. For concreteness, we restrict our analysis to the regime of Luttinger liquid parameters $K_{c2} \gtrsim 0.5$, and comment on the opposite regime in Sec.~\ref{sec:outlook}.

\subsubsection{Forward scattering}
In the case of forward scattering, the arguments in the sine-Gordon term in Eq.~\eqref{eq:hsoi} read
\begin{subequations}
\begin{align}
\Psi_1^f &= \frac{\phi_{1c}-\phi_{1s}+\theta_{1c}-\theta_{1s}-\phi_{2c}-\phi_{2s}-\theta_{2c}-\theta_{2s}}{\sqrt{2}}~,\\
\Psi_2^f &= \frac{\phi_{1c}+\phi_{1s}-\theta_{1c}-\theta_{1s}-\phi_{2c}+\phi_{2s}+\theta_{2c}-\theta_{2s}}{\sqrt{2}}~.
\end{align}\label{eq:arg1}
\end{subequations}
While $\Psi_{1}^f$ and $\Psi_2^f$ commute with each other, they do not commute with themselves, $[\Psi_{k}^f(x),\Psi_{l}^f(x')]=2i\pi\delta_{kl} {\rm sgn}(x'-x)$. This non-trivial commutation relation encodes that although forward scattering lifts the inter-subband crossings, it cannot open up a gap.

\subsubsection{Backscattering}
For backscattering, which occurs in the situation depicted in Fig.~\ref{fig:scatterings}$(c)$, the arguments of the sine-Gordon term in Eq.~\eqref{eq:hsoi} are given by
\begin{subequations}
\begin{align}
\Psi_1^b &= \frac{\phi_{1c}-\phi_{1s}+\theta_{1c}-\theta_{1s}+\phi_{2c}+\phi_{2s}-\theta_{2c}-\theta_{2s}}{\sqrt{2}}~,\\
\Psi_2^b &= \frac{\phi_{1c}+\phi_{1s}-\theta_{1c}-\theta_{1s}+\phi_{2c}-\phi_{2s}+\theta_{2c}-\theta_{2s}}{\sqrt{2}}~.
\end{align}\label{eq:arg2}
\end{subequations}
Since backscattering does open up a (partial) gap in the spectrum, the fields $\Psi_{1}^b$ and $\Psi_2^b$ commute both with themselves and with each other.

\begin{figure}
\centering
\includegraphics[scale=0.85]{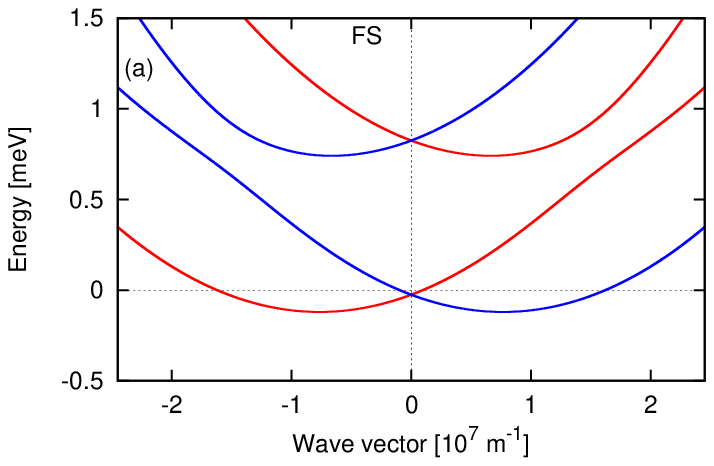}\\
\includegraphics[scale=0.85]{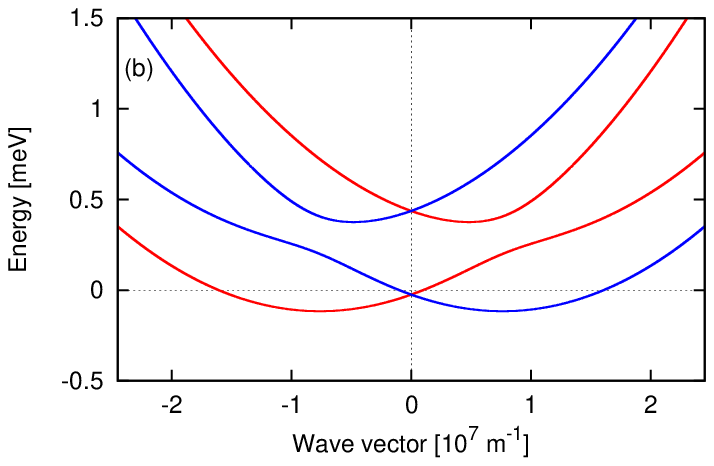}\\
\includegraphics[scale=0.85]{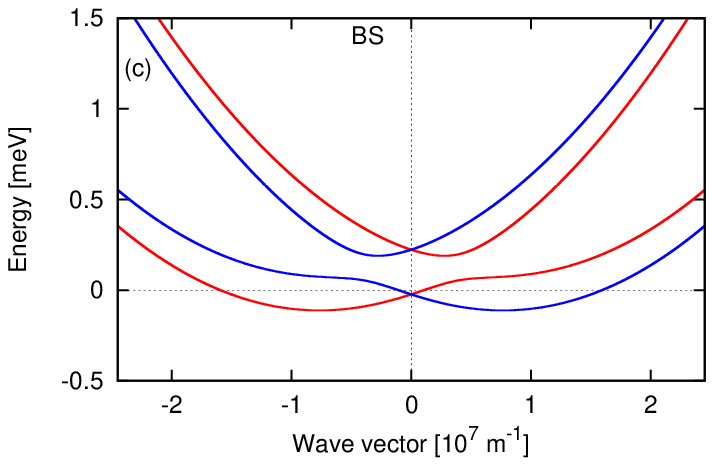}
\caption{Bandstructure of the two-subband quantum wire in the non-interacting case for relatively large interband couplings (all parameters are chosen as in Fig.~\ref{fig:scatterings}, except for $\alpha+\beta = 0.5\,\alpha_0$). The band offsets are again given by $\delta_{12} = 0.8\,\text{meV}$ in panel $(a)$, $\delta_{12} \approx 0.4\,\text{meV}$ in panel $(b)$, and $\delta_{12} = 0.2\,\text{meV}$ in panel $(c)$.} 
\label{fig:scatterings2}
\end{figure}

\section{inter-subband anticrossings and their renormalizations}\label{sec:rg}
The inter-subband spin-orbit coupling giving rise to the sine-Gordon terms in Eq.~\eqref{eq:hsoi} lifts the band crossings shown in Fig.~\ref{fig:scatterings}. In Fig.~\ref{fig:scatterings2}, we plot the bandstructures corresponding to $\alpha - \beta = \alpha_0$, and an interband coupling $\alpha+\beta = 0.5\,\alpha_0$, while all other parameters are chosen as in Fig.~\ref{fig:scatterings}. In particular, we stick to the non-interacting case $U=0$ for this figure, such that the fermionic Hamiltonian given in Eq.~\eqref{eq:Ham} can be diagonalized. In the remainder, we discuss how the spectra shown in Fig.~\ref{fig:scatterings2} are renormalized by electron-electron interactions as a function of the inter-subband spacing $\delta_{12}$, or the intra-subband spin-orbit coupling $\alpha-\beta$, which are experimentally tunable by electrostatic gates.

\begin{figure}
\centering
\includegraphics[scale=0.85]{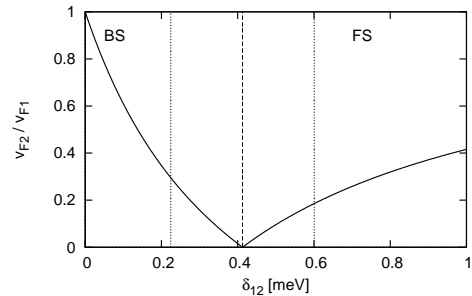}
\caption{Ratio of the Fermi velocities ${v_{F2}}/{v_{F1}}$ at the crossing points (see Fig.~\ref{fig:scatterings}). The dashed vertical line indicates the interband offset at which the band crossings occur at the bottom of the second band. For smaller $\delta_{12}$, the inter-subband spin-orbit coupling gives rise to backscattering (BS), while larger values of $\delta_{12}$ lead to forward scattering (FS). The dotted lines delimit the excluded regime in which strong interactions in the second subband eventually modify the physics (we choose $K_{2c} = 0.5$ as the criterion to distinguish weak from strong interactions).} 
\label{fig:velocities}
\end{figure} 

\subsection{Perturbative RG analysis}
\begin{figure}
\centering
\includegraphics[scale=0.85]{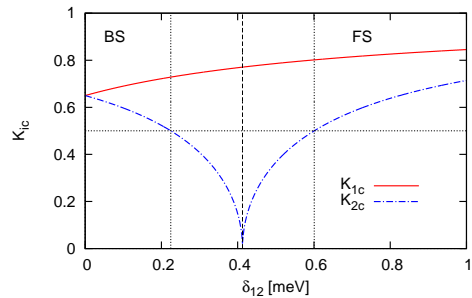}
\caption{Luttinger liquid parameters $K_{ic}$ in the charge sector of band $i = 1,2$ if the chemical potential intersects the crossing points of the spectra shown in Fig.~\ref{fig:scatterings}, as a function of the band offset $\delta_{12}$. For different band offsets, the crossing point occurs at different points of the bandstructure, which alters the Fermi velocities. This in turn modifies the value of the Luttinger liquid parameters. Like in Fig.~\ref{fig:velocities}, the dotted vertical lines (and the dotted horizontal line) indicate $K_{2c}=0.5$, the dashed vertical line indicates the transition point between forward (FS) and backscattering (BS).} 
\label{fig:k1ck2c}
\end{figure}

The renormalization of the inter-subband spin-orbit couplings due to the Coulomb repulsion can be addressed with a Luttinger liquid renormalization group (RG) analysis\cite{giamarchi_book} of the sine-Gordon terms given in Eq.~\eqref{eq:hsoi}. The strength of the renormalizations depends on the ratio of kinetic energy and Coulomb repulsion. As can be inferred from Eq.~\eqref{eq:2band_ham_simpl}, a modification of either the band offset $\delta_{12}$ or the intra-subband spin-orbit coupling $(\alpha-\beta)$ not only allows to change between forward and backscattering, but also leads to modified values of the Fermi velocities at the crossing point,
\begin{align}
\frac{v_{F2}}{v_{F1}}&=\frac{\left|1-\frac{\delta_{12}}{2 m(\alpha-\beta)^2}\right|}{1+\frac{\delta_{12}}{2 m(\alpha-\beta)^2}}~,\label{eq:vel_rat}
\end{align}
which is depicted in Fig.~\ref{fig:velocities}. The modified Fermi velocities in turn affect the Luttinger liquid parameters $K_{ic} = [1+2U/(\pi v_{Fi})]^{-1/2}$ as depicted in Fig.~\ref{fig:k1ck2c}, and thus the RG flow.

The RG equations for the inter-subband spin-orbit couplings $\alpha_{12}^{(i)}$ are derived in a real space RG scheme with a running short-distance cutoff $a(b) = a_0\,b$ by expanding the action corresponding to Eq.~$\eqref{eq:ll_ham}$ to first order in $\alpha_{12}^{(i)}$, see Appendix \ref{append:12trafo}. Taking into account that the argument of the sine-Gordon term changes between forward scattering and backscattering, we obtain the RG equation

\begin{align}
\frac{d \alpha_{12}^{(i)}}{d\ln(b)} = (1-g_{12})\,\alpha_{12}^{(i)}~,
\end{align}
where the RG scaling dimension $g_{12}$ depends on the Luttinger liquid parameters set by the Coulomb interaction strength, and the velocities. Its calculation requires the diagonalization of the quadratic part of the bosonized Hamiltonian, which is detailed in Appendix \ref{append:12trafo}. Fig.~\ref{fig:scaling_dim} shows $(1-g_{12})$ as a function of the band  offset $\delta_{12}$.

\begin{figure}
\centering
\includegraphics[scale=0.85]{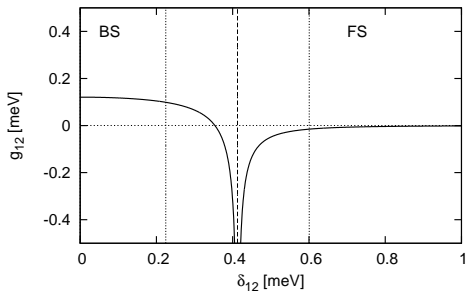}
\caption{The parameter ($1-g_{12}$) as a function of the band offset $\delta_{12}$ ($g_{12}$ is the RG scaling dimension of the inter-subband spin-orbit couplings). As in Fig.~\ref{fig:velocities}, the vertical dashed line  marks the transition between forward (FS) and backscattering (BS), the vertical dotted  lines indicate the regime in which the band crossing occurs close to the bottom of the upper bands.} 
\label{fig:scaling_dim}
\end{figure} 
We find that $(1-g_{12})$ is always negative for forward scattering, resulting in an effectively reduced interband spin-orbit coupling. Backscattering, on the other hand, is generally enhanced under RG. The scaling dimension $g_{12}$, finally, grows (and, extrapolating our theory, is even pushed to RG irrelevant values $g_{12} >2$) when the band crossings approach the bottom of the upper bands. This trend can be traced back to the ratio of Fermi velocities shown in Fig.~\ref{fig:velocities}. The more these Fermi velocities differ, the harder it is to scatter, say, a fast electron from band one into the slowly moving modes of band two. This effect is known from Coulomb-drag setups,\cite{drag_literature1,drag_literature2,drag_literature3} and has also been analyzed for the RKKY interaction in two-subband quantum wires.\cite{meng_2bands_13} 

\subsection{Renormalized anticrossing gaps in the backscattering regime}
\begin{figure}
\centering
\includegraphics[scale=0.85]{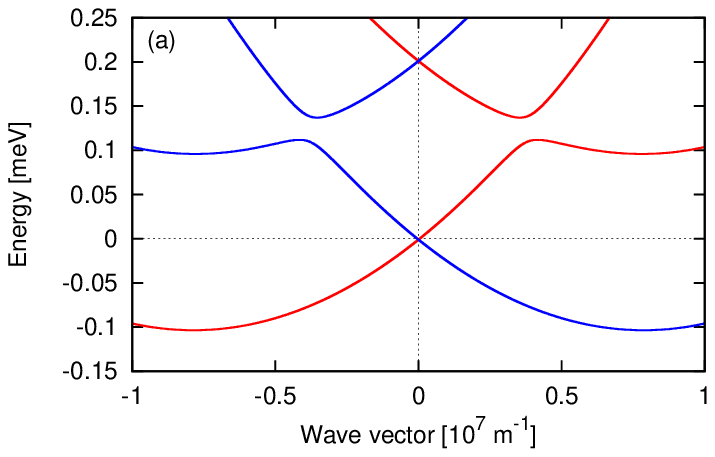}\\
\includegraphics[scale=0.85]{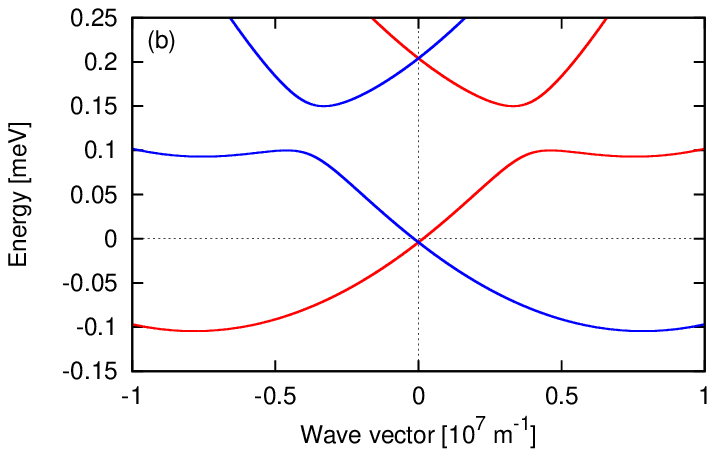}
\caption{Bandstructure of the two subband quantum wire for a band offset of $\delta_{12} = 0.1 \, \text{meV}$, $\alpha-\beta = \alpha_0$, and $\alpha+\beta = 0.1\,\alpha_0$. Panel $(a)$ shows the unrenormalized bandstructure, panel $(b)$ the one renormalized by electron-electron interactions.}
\label{fig:scatterings3}
\end{figure}

We now turn to the case where inter-subband spin-orbit coupling allows for backscattering, such that it results in a partial gap. Assuming both finite size effects and finite temperature effects to be small, the RG flow is integrated until the running gap associated with the inter-subband coupling reaches the running band width (at high temperatures or in short wires, the RG flow is cut off by either the temperature or the wire length, respectively, and renormalization effects due to Coulomb interactions are less pronounced). The gap can be defined by the expansion of the cosine to second order.\cite{giamarchi_book} At the end of the RG flow, the renormalized inter-subband spin-orbit coupling $\alpha_{12}^{(i)}{}^*$  reads
\begin{align}
\alpha_{12}^{(i)}{}^* = \alpha_ {12}^{(i)}\,b^*{}^{1-g_{12}} = \frac{v_{F2}}{a}\,\left(\frac{a \alpha_{12}^{(i)}}{v_{F2}}\right)^{1/(2-g_{12})}~,\label{eq:scaling_dim}
\end{align}
where $b^*{}^{2-g_{12}} = v_{F2}/({a\alpha_{12}^{(i)}})$ is the RG scale at which the flow stops. The gap therefore scales as an interaction-dependent power law of the bare inter-subband spin-orbit coupling, $\alpha_{12}^{(i)}{}^{1/(2-g_{12})}$, which only reduces to a linear dependence in the non-interacting case. We illustrate the renormalization for a band offset of $\delta_{12} = 0.2\,\text{meV}$, for a bare spin-orbit coupling of $\alpha+\beta = 0.1\,\alpha_0$ in Fig.~\ref{fig:scatterings3}, by feeding the renormalized values $\alpha_{12}^{(1)}{}^*$ and $\alpha_{12}^{(2)}{}^*$ back into the non-interacting theory, and find that the gap of the anticrossings is renormalized to roughly twice its bare value for the depicted set of parameters. The fact that interactions do not affect the anticrossings more strongly has two reasons. The value of $1-g_{12}$ is relatively small, and the bare couplings $\alpha_{12}^{(i)}$ are large (the relative shift of the dispersions for spin up and spin down electrons in momentum space is of the order of the Fermi momentum). The RG flow is thus rapidly cut off by the gap.

\section{Closing of the anticrossing by electric fields}\label{sec:closing}
The renormalization of the inter-subband anticrossings becomes especially important when the system is  tuned to the point of equally strong Rashba and Dresselhaus spin orbit interactions, $\alpha\to-\beta$, where the anticrossing splittings close. At this special point, the system exhibits a number of unusual properties, such as a negative magnetoresistance (weak localization), \cite{pikus_95} and giant spin relaxation anisotropies.\cite{aserkiev_99} It furthermore allows for the construction of nonballistic spin-field-effect transistors,\cite{loss_equal} and for a persistent spin helix,\cite{loss_equal,bernevig_equal,koralek_spin_helix_09,duckheim_09}, the control of spin decoherence in quantum dots, \cite{golovach_04,bulaev_05,golovach_08} and it guarantees the absence of spontaneous magnetic order.\cite{loss_11} These effects are intimately related to the conservation of the spin along the axis set by the intersubband spin-orbit coupling for $\alpha\to-\beta$.\cite{loss_equal,bernevig_equal} Since the intra-subband spin-orbit coupling can furthermore be removed by a gauge transformation (in the absence of a Zeeman term),\cite{braunecker_prb_10} the 
special point $\alpha=-\beta$ can also be understood as a wire without spin-orbit coupling.

\begin{figure}
\centering
\includegraphics[scale=0.85]{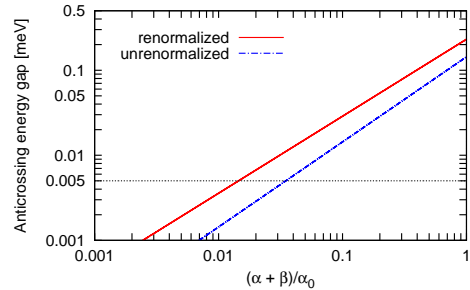}
\caption{Scaling of the inter-subband anticrossings as a function of the bare inter-subband interaction $\alpha+\beta$ for fixed $\beta = -\alpha_0/2$. The unrenormalized gap depends linearly on $\alpha+\beta$, while the renomormalized one obeys the power law defined in Eq.~\eqref{eq:scaling_dim}. As an example, this graphs contrasts the renormalized and unrenormalized anticrossing gaps for $\delta_{12} = 0.2$, in which case the anticrossing gap scales with the exponent $1/(2-g_{12})\approx 0.91$ as function of the bare inter-subband spin-orbit coupling. The dotted line marks an energy of $0.005\,\text{meV} \equiv 50\,\text{mK}$, which constitutes a realistic experimental measurement resolution.}
\label{fig:gap_scaling}
\end{figure}

 The size of the renormalized inter-subband anticrossings during the tuning of $\alpha+\beta$ to zero can be monitored in situ, for instance by virtue of tunneling spectroscopy,\cite{spectroscopy_02,auslaender_spin_charge_seperation_exp_05,jompol_kc_ks} or by optical techniques.\cite{kim_96,segovia_99,kim_arpes_05} Its non-linear dependence on the bare parameter $\alpha+\beta$ as a function of the electric field is a direct measure of Luttinger liquid physics, and more precisely of the RG scaling dimension $g_{12}$. Figure~\ref{fig:gap_scaling} displays the nonlinear scaling of the interband anticrossings when $\alpha \to -\beta$ for fixed $\beta=-\alpha_0/2$ and $\delta_{12}=0.2$, that is in the backscattering regime, where the inter-subband spin-orbit coupling is enhanced by electron-electron interactions. We find that the Luttinger liquid power law can be measured over several decades (especially since the bare value of $\alpha+\beta$ can be tuned to exceed $\alpha_0$), even when taking into account realistic energy 
resolutions of $0.005\,\text{meV}$, or equivalently temperatures up to $\sim 50\,\text{mK}$, or finite size effect for wire lengths of a few microns. While the variation of $\alpha$ at fixed $\beta$ implies a variation of the intra-subband spin-orbit coupling $\alpha-\beta$, Fig.~\ref{fig:gap_scaling} shows that this does not affect the power law of the anticrossing gap in a noticeable way. The measurement range is, in our model, limited by the subband spacing. When the size of the anticrossing splitting is of the order of the spacing to the next higher subbands, the two-subband model used here becomes inaccurate. To remedy this shortcoming, one may increase the subband spacing $\delta_{12}$, and the intra-subband spin-orbit coupling $\alpha-\beta$ in such a way that the anticrossings still occur between modes moving in opposite directions. This allows to increase the maximal value of the inter-subband spin-orbit coupling $\alpha+\beta$ whilst keeping the two-subband approximation justified, and to thereby 
extend the measurement range described by the present theory.

\section{Band crossings at the bottom of the upper bands}\label{sec:outlook}
As mentioned above, the regime of anticrossings occurring at the bottom of the upper bands is not captured by our theory: not only the ratio of Coulomb repulsion energy to kinetic energy becomes very large, but also the curvature of the upper bands becomes eventually important. These effects are beyond our theory based on a Luttinger liquid approach for weakly interacting electrons. We can, however, expect some of the features of our calculation to remain valid even for $K_{c2} < 0.5$, a regime which can only be reached with long-range interactions. Most importantly, we expect the RG scaling dimension $g_{12}$ to grow when band bottoms of the upper bands are approached. One can consequently expect the size of the inter-subband anticrossing gap to be largely reduced  in the strongly interacting regime close to the band bottoms. It would thus be worthwhile to also monitor the size of the inter-subband anticrossing gaps as the chemical potential approaches the bottom of the upper bands in an experiment.

\section{Effects of an applied magnetic field}\label{sec:outlook2}
So far, we have not included the effects of a possibly applied magnetic field. Depending on its direction, such a field affects the properties of the wire in a number of ways. We discuss the two limiting cases of a field aligned either parallely or perpendicularly to the spin quantization axis set by the intra-subband spin-orbit coupling. For a general direction, the magnetic field exhibits all of the effects discussed in Secs.~\ref{subsec:mag1} and \ref{subsec:mag2}.

\subsection{Magnetic field parallel to the spin-orbit direction}\label{subsec:mag1}
\begin{figure}
\centering
\includegraphics[scale=0.85]{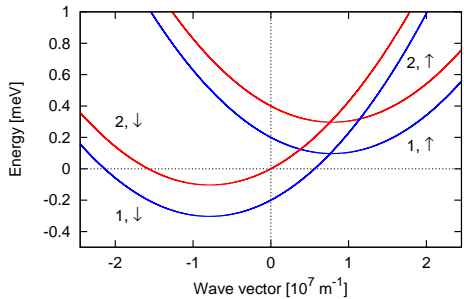}
\caption{Band structure of the two subband quantum wire in the non-interacting case, without interband couplings, and with an applied magnetic field parallel to the spin quantization axis set by the intra-subband spin-orbit coupling (we use $U = 0$, $\alpha+\beta = 0$, and $\alpha-\beta = \alpha_0$ for $\delta_{12}=0.2\,\text{meV}$ and a Zeeman energy of $\pm0.2\,\text{meV}$ for spin up and spin down). For this value of the magnetic field, crossing between the bands $1,\downarrow$ and $2,\uparrow$ corresponds to forward scattering, while the one between $1,\uparrow$ and $2,\downarrow$ corresponds to backscattering.
} 
\label{fig:scatterings_b}
\end{figure}

If the magnetic field is applied along the spin quantization axis set by the intra-subband spin-orbit coupling, its Zeeman effect shifts spin up and down relative to each other in energy. The initially degenerate anticrossings between the bands $1,\downarrow$ and $2,\uparrow$, and $1,\uparrow$ and $2,\downarrow$ then occur at different energies, see Fig.~\ref{fig:scatterings_b}. The stronger the magnetic field becomes, the more the inter-subband crossings (eventually lifted by the spin-orbit coupling) are pushed towards the forward scattering regime. Importantly, since the crossings are not degenerate in energy anymore, they are also not equally affected by the magnetic field. We find that if without magnetic field, the crossings occur in the backscattering regime, an intermediate magnetic field can create the interesting situation that one of the (eventually lifted) crossings corresponds to forward scattering, while the other one corresponds to backscattering. This is precisely the case in the situation depicted in Fig.~\ref{fig:scatterings_b}. As can be inferred from the discussion of Sec.~\ref{sec:rg}, electron-electron interactions then result in differently large renormalizations for the two anticrossings. Since, furthermore, the bare energy gaps of the anticrossings are identical, $\alpha_{12}^{(1)}=\alpha_{12}^{(2)}$, we thus find that any difference in gaps of the two anticrossings is a direct signature of electron-electron interactions, and their interplay with the magnetic field.

\subsection{Magnetic field perpendicular to the spin-orbit direction}\label{subsec:mag2}
\begin{figure}
\centering
\includegraphics[scale=0.85]{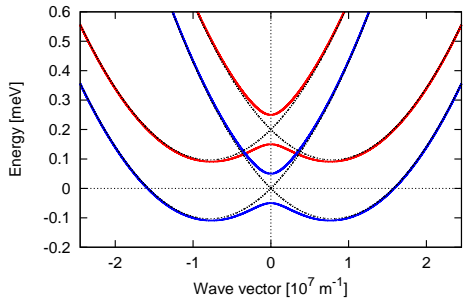}
\caption{Band structure of the two subband quantum wire in the non-interacting case, without interband couplings, and with an applied magnetic field perpendicular to the spin quantization axis set by the intra-subband spin-orbit coupling (we use $U = 0$, $\alpha+\beta = 0$, and $\alpha-\beta = \alpha_0$ for $\delta_{12}=0.2\,\text{meV}$ and $\Delta_{B}=0.05\,\text{meV}$). The dotted lines indicate the bands without magnetic field. 
} 
\label{fig:scatterings_b_2}
\end{figure}

As is well-known from the single subband case, a magnetic field applied perpendicular to the direction set by the intra-subband spin-orbit coupling opens partial gaps at the crossings of the bands $i,\uparrow$ and $i,\downarrow$ (with $1=1,2$), see Fig.~\ref{fig:scatterings_b_2}. At the inter-subband crossing points, the distinct inversion symmetries of the two transversal wave functions prevents the opening of a gap in the absence of spin-orbit coupling. If the chemical potential is tuned to either of the crossing points between the subband $i,\uparrow$ and $i,\downarrow$, the magnetic field is described by the Hamiltonian

\begin{align}
H_B^{(i)}=\int dx\,\frac{\Delta_{B}}{\pi a}\,\cos\left(\sqrt{2}(\phi_{ic}+\theta_{is})\right)~,
\end{align}
where $\Delta_{B}=g\mu_B B/2$ ($g$ is the $g$-factor, $\mu_B$ denotes Bohr's magneton, while $B$ is amplitude of the applied magnetic field). Away from these crossing points, the Hamiltonian describing the magnetic field contains rapidly oscillating factors, and can thus be dropped.  In complete analogy to the single subband case,\cite{braunecker_prb_10} electron-electron interactions renormalize the partial gaps opened by the magnetic field to the values 

\begin{align}
\Delta_i^* = \Delta_B \left(\frac{\hbar v_{F,i}}{a_0\Delta_B}\right)^{(1-g_{\Delta,i})/(2-g_{\Delta,i})}~,
\end{align}
where the scaling dimensions $g_{\Delta,i}$ depend on the strength of electron-electron interactions, and the Fermi velocities within the bands, as well as on the subband mixing due to both inter-subband spin-orbit coupling, and Coulomb repulsion. If the subband offset $\delta_{12}$ is larger than the energy scale associated with the intra-subband spin-orbit coupling, such that the second subband lives at energies higher than the energy of the crossing between the bands $1,\uparrow$ and $1,\downarrow$, we find that $g_{\Delta,1} = (K_{1c}+1/K_{1s})/2$.  Considering concretely the Luttinger liquid parameters $K_{ic}=0.65$ and $K_{is}=1$,\cite{spectroscopy_02,auslaender_spin_charge_seperation_exp_05,jompol_kc_ks} this exponent becomes $g_{\Delta,1}\approx 0.85$, and thus significantly different from its noninteracting value is $g_{\Delta,1}^{(0)}=1$. It is, however, of the same order as the scaling exponent $g_{12}$ of the inter-subband anticrossings in the backscattering regime, as can be inferred from Fig.~\ref{fig:scaling_dim}. The calculation of the scaling dimension $g_{\Delta,2}$ (and similarly of $g_{\Delta,1}$ for small band offsets $\delta_{12}$) requires the diagonalization of the Hamiltonian according to the discussion of Appendix \ref{append:12trafo}. Because the power law behavior of $\Delta_{B,1}$ and $\Delta_{B,2}$ can be measured over several decades by simply tuning the external magnetic field, we conclude in analogy to Sec.~\ref{sec:closing} that the monitoring of the gaps as a function of the applied field $B$ constitutes an additional signature of Luttinger liquid physics. As a further effect of the magnetic field, the distortion of the bandstructure due to the presence of additional gaps as compared to the case without field modifies the Fermi velocities at the eventually lifted inter-subband crossing points, and therefore also modifies the renormalization of the inter-subband anticrossings.

\section{Transport signatures}\label{sec:trans}
In electronic transport, the opening of partial gaps is heralded by a reduction of the conductance. This reduction can be calculated in an inhomogenous Luttinger model, which also takes into account the leads. \cite{conductance_ll_leads1,conductance_ll_leads2,conductance_ll_leads3,meng_fract} A universally quantized conductance is then obtained in the scaling limit of an RG relevant sine-Gordon potential, such as the inter-subband spin-orbit coupling, or a magnetic field perpendicular to the spin-quantization axis set by the intra-subband spin-orbit coupling. Electron-electron interactions are a crucial ingredient in observing the conduction reduction in an experiment,\cite{braunecker_prb_10,meng_fract,scheller_13} since they drive the system towards the universal scaling limit. Put differently, interactions boost the magnitude of the gap as compared to, for instance, temperature, and finite size gaps, which reduces non-universal corrections to the conductance present for small gaps.

In a quantum wire with two subbands, and both Rashba and Dresselhaus spin-orbit couplings, we thus find that the conductance is very sensitive to the magnitude, and orientation of an external electric field. If the field is close to the point of partial compensation between Rashba and Dresselhaus couplings, $\alpha\approx-\beta$, the conductance through the ballistic wire amounts to $G=4\,e^2/h$. If the electric field deviates from this special point, the anticrossing gaps open, and are quickly renormalized to stronger values by electron-electron interactions. This causes the conductance to drop to half its value, i.e.~$G\to 2\,e^2/h$.

\section{Summary}
In this work, we have analyzed the effect of electron-electron interactions on spin-orbit couplings in multi-subband quantum wires. Depending on the ratio of the energy spacing between the subbands and the intra-subband spin-orbit coupling, electron-electron interactions can either increase or decrease the inter-subband spin-orbit coupling. For large spacings or small Rashba interactions, when the inter-subband spin-orbit interaction couples electrons moving into the same direction, the inter-subband spin-orbit coupling lifts the degeneracy between the bands, but does not open up a gap. If spin-orbit interaction couples electrons moving into opposite directions, a gap opens. Unless the chemical potential is close to the bottom of the upper bands, this gap is enlarged by the presence of electron-electron interactions. Analyzing this renormalization with a renormalization group approach, we showed that the effective spin-orbit couplings are interaction-dependent power laws of the bare spin-orbit parameters, and thus of an applied electric field. 

We have discussed how the scaling dimension of the inter-subband spin-orbit coupling can be measured by monitoring the closing of the anticrossings as one fine-tunes the bare parameters to the point of partial compensation between Rashba and Dresselhaus couplings. We then commented on the strongly interacting regime occurring when the chemical potential is close to the bottoms of the upper bands, where our calculation hints at a strong reduction of the anticrossing splitting. 

Finally, we discussed the effects of an external magnetic field. If the latter is applied parallel to the spin quantization axis set by the intra-subband spin-orbit coupling, a suitably chosen field can shift the inter-subband crossings such that one of them results in forward scattering, while the other one corresponds to backscattering. In this case, electron-electron interactions enlarge the anticrossing that results from backscattering, while they decrease the one due to forward scattering. The difference in the gap size of the anticrossings is thus a convenient signature of electron-electron interactions. If the magnetic field is applied perpendicular to the direction set by the intra-subband spin-orbit coupling, it opens additional partial gaps. Similar to the inter-subband anticrossings, these gaps experience a substantial renormalization in the presence of electron-electron interactions. Monitoring the partial gaps opened by the magnetic field as a function of the strength of this field thus again allows one to observe the scaling dimension of the magnetic field in an experiment, which constitutes an further direct measure of Luttinger liquid physics. The distortions of the bandstructure resulting from the gaps opened by the magnetic field furthermore affect the Fermi velocities at the (lifted) crossing points, and therefore changes the renormalization of the anticrossings. Finally, we have discussed that the conductance through the wire is reduced by a factor of two when the inter-subband anticrossings open, and get renormalized due to electron-electron interactions.

In a future work, it would not only be interesting to investigate the physics close to the bottoms of the upper bands in more detail, but also to analyze the effects of disorder, which we have neglected in the present work. In general, a quantum wire with scalar impurities is known to be susceptible to Anderson localization. One can speculate, however, that localization is suppressed in the quasi-helical regime just below the bottom of the second band, similar to the single subband case with applied magnetic field.\cite{braunecker_prb_13}

\acknowledgements
This work has been supported by the Swiss NF, NCCR QSIT, and the Harvard Quantum Optical Center (JK).

\appendix
\section{Diagonalization of the quadratic part of the Hamiltonian}\label{append:12trafo}
The diagonal electronic Hamiltonian is  obtained from $H_1+H_2$ [see Eq. (\ref{eq:ll_ham})] by the canonical transformation
\begin{subequations}\label{eq:diagonalization_trafo}
 \begin{align}
\phi_{1c}&= \sqrt{\frac{u_{1c}K_{1c}}{u_{c+}\,(1+\mathcal{A}_c^2)}}\,\phi_{c+}+\sqrt{\frac{\mathcal{A}_c^2\,u_{1c}K_{1c}}{u_{c-}(1+\mathcal{A}_c^2)}}\,\phi_{c-}~,\label{eq:diagonalization_trafo_pc1}\\
\phi_{2c}&= \sqrt{\frac{\mathcal{A}_c^2\,u_{2c}K_{2c}}{u_{c+}(1+\mathcal{A}_c^2)}}\,\phi_{c+}- \sqrt{\frac{u_{2c}K_{2c}}{u_{c-}(1+\mathcal{A}_c^2)}}\,\phi_{c-}~,\\
\theta_{1c}&=\sqrt{\frac{u_{c+}}{u_{1c}K_{1c}(1+\mathcal{A}_c^2)}}\,\theta_{c+} + \sqrt{\frac{\mathcal{A}_c^2\,u_{c-}}{u_{1c}K_{1c}(1+\mathcal{A}_c^2)}}\,\theta_{c-}~, \\
\theta_{2c}&=\sqrt{\frac{\mathcal{A}_c^2\,u_{c+}}{u_{2c}K_{2c}(1+\mathcal{A}_c^2)}}\,\theta_{c+} - \sqrt{\frac{u_{c-}}{u_{2c}K_{2c}(1+\mathcal{A}_c^2)}}\,\theta_{c-} ~,
 \end{align}
\end{subequations}
with the velocities

\begin{align}\label{eq:mixed_velocities}
 &u_{c\pm} =\\
 & \sqrt{\frac{u_{1c}^2+u_{2c}^2}{2}\pm\sqrt{\left(\frac{u_{1c}^2-u_{2c}^2}{2}\right)^2+U_{12}^2 u_{1c}K_{1c}u_{2c}K_{2c}}}~,\nonumber
\end{align}
and with
\begin{align}\label{eq:mixing_coeff_coulomb}
 \mathcal{A}_c= \frac{2U_{12}\,\sqrt{u_{1c}K_{1c}u_{2c}K_{2c}}}{\sqrt{\left(u_{1c}^2-u_{2c}^2\right)^2+4 U_{12}^2\,u_{1c}K_{1c}u_{2c}K_{2c}}+u_{1c}^2-u_{2c}^2} ~.
\end{align}
Using $K_{is} = 1$, the quadratic part of the electronic Hamiltonian can then be written as

\begin{align}\label{eq:diag_el_ham}
 H_{\rm e} &= \sum_{k=\pm}\frac{u_{ck}}{2\pi}\int dx' \,\left( \left(\partial_z\phi_{ck}\right)^2 + \left(\partial_z\theta_{ck}\right)^2\right)\\
  &+\sum_{i=1,2}\frac{v_{Fi}}{2\pi}\int dx' \,\left( \left(\partial_z\phi_{is}\right)^2 + \left(\partial_z\theta_{is}\right)^2\right).\nonumber
\end{align}
A generalized form of the transformation given in Eq.~\eqref{eq:diagonalization_trafo} would allow one to take into account the spin density-density interaction, charge current-current interaction and spin current-current interaction neglected here. In order to calculate the scaling dimension of the sine-Gordon terms given in Eq.~\eqref{eq:hsoi}, one first needs to decompose the fields $\Psi_{1,2}^{f,b}$ into the new fields $\phi_\pm$ and $\theta_\pm$. Considering first forward scattering, we find
\begin{align}
\Psi_{1}^f&=\sum_{i=\pm}(c_i\phi_i+d_i\theta_i)-\sum_{j=1,2}\frac{\phi_{js}+\theta_{js}}{\sqrt{2}}\nonumber\\
\Psi_{2}^f&=\sum_{i=\pm}(c_i\phi_i-d_i\theta_i)+\sum_{j=1,2}\frac{\phi_{js}-\theta_{js}}{\sqrt{2}}~,\nonumber
\end{align}
where the constants $c_i$ and $d_i$ follow from Eqs.~\eqref{eq:diagonalization_trafo}, and read
\begin{subequations}
\begin{align}
c_+ &= \sqrt{\frac{u_{1c}K_{1c}}{2u_{c+}\,(1+\mathcal{A}_c^2)}}-\sqrt{\frac{\mathcal{A}_c^2\,u_{2c}K_{2c}}{2u_{c+}(1+\mathcal{A}_c^2)}}~,\\
d_+ &=\sqrt{\frac{u_{c+}}{2u_{1c}K_{1c}(1+\mathcal{A}_c^2)}}-\sqrt{\frac{\mathcal{A}_c^2\,u_{c+}}{2u_{2c}K_{2c}(1+\mathcal{A}_c^2)}}~,\\
c_- &=\sqrt{\frac{\mathcal{A}_c^2\,u_{1c}K_{1c}}{2u_{c-}(1+\mathcal{A}_c^2)}}+\sqrt{\frac{u_{2c}K_{2c}}{2u_{c-}(1+\mathcal{A}_c^2)}}~,\\
d_- &=\sqrt{\frac{\mathcal{A}_c^2\,u_{c-}}{2u_{1c}K_{1c}(1+\mathcal{A}_c^2)}}+\sqrt{\frac{u_{c-}}{2u_{2c}K_{2c}(1+\mathcal{A}_c^2)}}~.
\end{align}
\end{subequations}

Since the quadratic part of the Hamiltonian is now diagonal, the scaling dimensions can be obtained from a standard RG analysis of the sine-Gordon potential.\cite{giamarchi_book} We find that $\alpha_{12}^{(1)}$ and $\alpha_{12}^{(2)}$ obey the same RG equation, 
\begin{align}
\frac{d \alpha_{12}^{(i)}}{d\ln(b)} = (1-g_{12})\,\alpha_{12}^{(i)}~,
\end{align}
with
\begin{align}
 g_{12}&=\sum_{i=\pm}\frac{c_i^2+d_i^2}{4}+\frac{K_{1s}+1/K_{1s}+K_{2s}+1/K_{2s}}{8}
\end{align}
for forward scattering. 

In the case of backscattering, on the other hand, we obtain
\begin{align}
\Psi_{1}^b&=\sum_{i=\pm}(\tilde{c}_i\phi_i+\tilde{d}_i\theta_i)-\frac{\phi_{1s}+\theta_{1s}-\phi_{2s}+\theta_{2s}}{\sqrt{2}}~,\nonumber\\
\Psi_{2}^b&=\sum_{i=\pm}(\tilde{c}_i\phi_i-\tilde{d}_i\theta_i)-\frac{\phi_{2s}+\theta_{2s}-\phi_{1s}+\theta_{1s}}{\sqrt{2}}~,\nonumber
\end{align}
with
\begin{subequations}
\begin{align}
\tilde{c}_+ &= \sqrt{\frac{u_{1c}K_{1c}}{2u_{c+}\,(1+\mathcal{A}_c^2)}}+\sqrt{\frac{\mathcal{A}_c^2\,u_{2c}K_{2c}}{2u_{c+}(1+\mathcal{A}_c^2)}}~,\\
\tilde{d}_+ &=\sqrt{\frac{u_{c+}}{2u_{1c}K_{1c}(1+\mathcal{A}_c^2)}}-\sqrt{\frac{\mathcal{A}_c^2\,u_{c+}}{2u_{2c}K_{2c}(1+\mathcal{A}_c^2)}}~,\\
\tilde{c}_- &=\sqrt{\frac{\mathcal{A}_c^2\,u_{1c}K_{1c}}{2u_{c-}(1+\mathcal{A}_c^2)}}-\sqrt{\frac{u_{2c}K_{2c}}{2u_{c-}(1+\mathcal{A}_c^2)}}~,\\
\tilde{d}_- &=\sqrt{\frac{\mathcal{A}_c^2\,u_{c-}}{2u_{1c}K_{1c}(1+\mathcal{A}_c^2)}}+\sqrt{\frac{u_{c-}}{2u_{2c}K_{2c}(1+\mathcal{A}_c^2)}}~.
\end{align}
\end{subequations}
In the backscattering regime, the RG flow of $\alpha_{12}^{(i)}$ is thus described by
\begin{align}
\frac{d \alpha_{12}^{(i)}}{d\ln(b)} = (1-g_{12})\,\alpha_{12}^{(i)}~,
\end{align}
with
\begin{align}
 g_{12}&=\sum_{i=\pm}\frac{\tilde{c}_i^2+\tilde{d}_i^2}{4}+\frac{K_{1s}+1/K_{1s}+K_{2s}+1/K_{2s}}{8}~.
\end{align}

%%%%%%%%%%%%%%%%%%%%%%%

\end{document}